\documentclass{article}

\usepackage{arxiv}

\usepackage[utf8]{inputenc} 
\usepackage[T1]{fontenc}    
\usepackage{hyperref}       
\usepackage{url}            
\usepackage{booktabs}       
\usepackage{amsfonts}       
\usepackage{nicefrac}       
\usepackage{microtype}      
\usepackage{lipsum}		
\usepackage{graphicx}
\usepackage{natbib}
\usepackage{doi}

\title{Bayesian Analysis of Extreme Precipitation Events and Forecasting Return Levels}

\author{ 
\hspace{1mm}Douglas E. Johnston\\
	Department of Applied Mathematics\\
	Farmingdale State College\\
	Farmingdale, NY 11735 \\
	\texttt{douglas.johnston@farmingdale.edu} \\
}

\begin{document}
\maketitle

\begin{abstract}
	In this study, we examine a Bayesian approach to analyze extreme daily rainfall amounts and forecast return-levels. Estimating the probability of occurrence and quantiles of future extreme events is important in many applications, including civil engineering and the design of public infrastructure.  In contrast to traditional analysis, which use point estimates to accomplish this goal, the Bayesian method utilizes the complete posterior density derived from the observations. The Bayesian approach offers the benefit of well defined credible (confidence) intervals, improved forecasting, and the ability to defend rigorous probabilistic assessments.  We illustrate the Bayesian approach using extreme precipitation data from Long Island, NY, USA and show that current return levels, or precipitation risk, may be understated.
\end{abstract}


\section{Introduction}
Extreme value theory (EVT) is the primary tool for the modeling and statistical analysis of extreme events and has found wide applications in fields as diverse as target detection \citep{broadwater2010}, communication systems \citep{ResRootz2000}, image analysis \citep{SJRoberts2000}, power systems \citep{shenoy2015}, and population studies \citep{Anderson3252}. That said, extreme value analysis (EVA) is of keen interest in the fields of finance \citep{dej_pmd_SPM2011} and climatology \citep{atg_cmc_2017}, where extreme events can have large, negative societal effects and impact public welfare significantly. Extreme climate events, such as extreme precipitation, have important implications in civil and infrastructure engineering where public infrastructure and human health are important considerations.

A critical step in analysing and forecasting future extreme events is determining the cumulative probability density curve from which projections of event probabilities can be made.  Often, it is desired to estimate event probabilities that would be considered rare and, to assist in this extrapolation process, historical data and parametric models are typically employed.  When the historical data corresponds to a series of maximum events (e.g., largest daily rainfall amounts in a year), EVT provides the desired form for the cumulative density in the generalized extreme value (GEV) distribution \citep{ekm_MEE_2003}. Fitting this parametric model to observable data then allows a more rigorous extrapolation for estimating extreme event probabilities.

From the cumulative density function (cdf), statistics of interest are estimated such as the $N$-year return-level, which is defined as the quantile corresponding to a probability of $1-1/N$.  For example, with annual extreme rainfall amounts, the $N$-year return-level is the daily rainfall amount such that the expected number of years to observe an annual maximum exceeding that level is $N$ years.  While this sounds a bit convoluted, it highlights that care must be taken when converting a statistic into a probabilistic statement \citep{coxnorthrop2002}. Typical return-levels of interest are $N = 25, 100, 500$ years which, given the limited historical records, results in a challenging problem.

Traditionally, estimates for the parameters of the GEV distribution are computed and these are, in turn, used to make inferences. Methods for estimating the GEV parameters include maximum likelihood and moment based methods \citep{coles_1999}, \citep{Nerantzaki_2022}. In addition, sensitivity analysis is generally employed based on the estimates having asymptotic properties, such as normality, along with other approximations.  While these assumptions may have validity, for large samples and well behaved, spherical distributions, their sole use in all situations is questionable \citep{coles_1996}.

The main purpose of this study is to highlight the benefits of a Bayesian approach to analyzing extreme events and apply this to climate data.  The Bayesian approach focuses on the complete posterior distribution, given the data, using it to compute estimates and confidence (credible) intervals as well as answer other probabilistic queries. The Bayesian method naturally incorporates recursion, when new data is available, it allows for cost functions in estimation and hypothesis testing, and it can make use of prior information, when available.  Outside of additional computer resources, there is seemingly little reason not to use a Bayesian methodology.

\begin{figure}
	\centering
    \includegraphics[width=.5\textwidth]{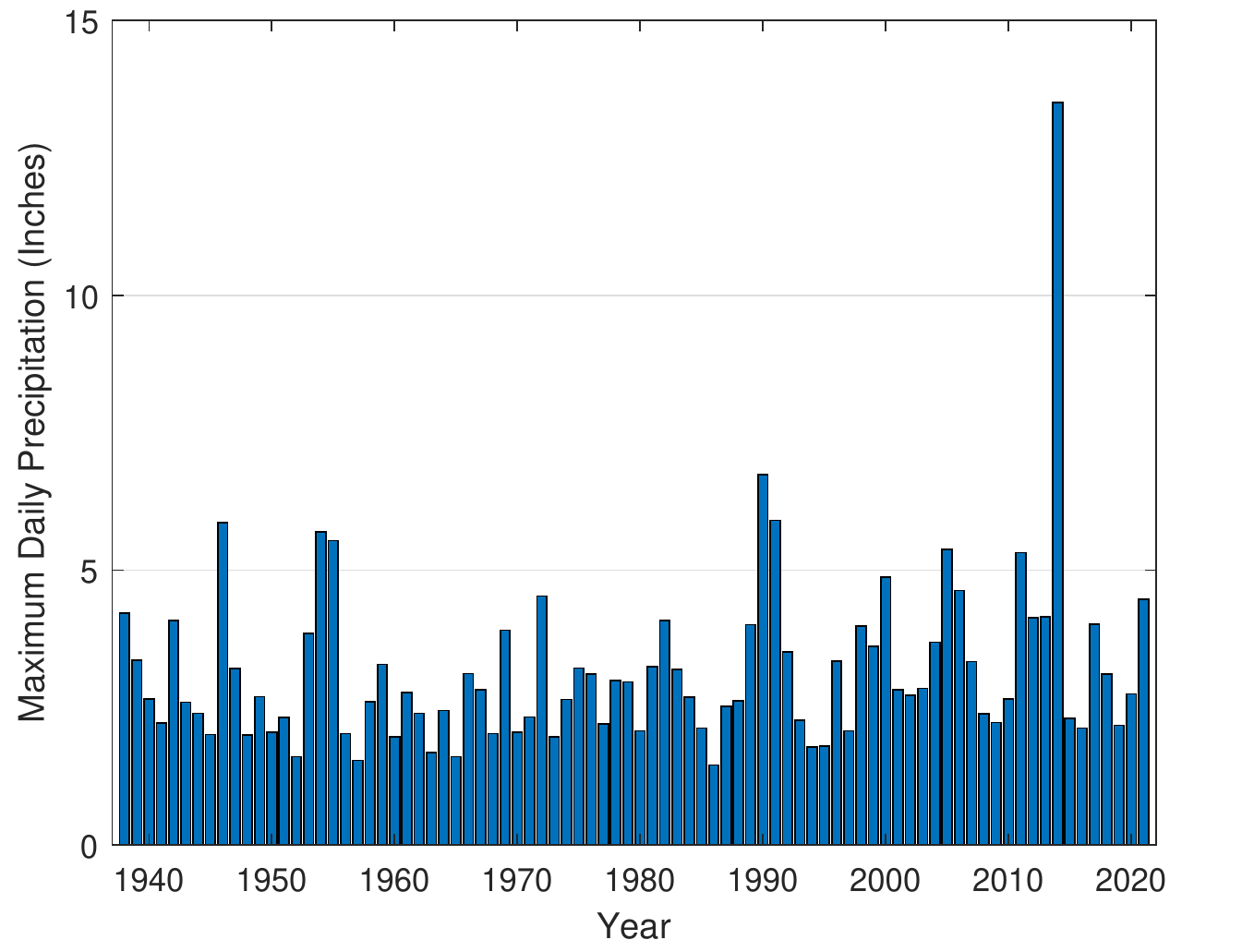}
    \caption{Annual block maximum of daily precipitation at ISP (1938 - 2021)}	\label{fig:fig1}
\end{figure}

To illustrate the Bayesian method, we analyzed annual maximums of daily precipitation. The data we used in this study were obtained from NOAA's National Centers for Environmental Information, Climate Data Online, website (https://www.ncdc.noaa.gov/cdo-web/).  Specifically, we obtained the maximum annual precipitation amount at the Islip LI MacArthur Aiport (ISP), New York, US weather station.  This data was available from 1964 to the present.  To increase the data used, we included earlier data from the nearby Patchogue weather station, which is about 5 miles away, going back to 1938.  The data is shown in Figure \ref{fig:fig1}. 

\section{Problem Formulation}
\label{sec:probform}

Let $y_k\in\mathbb{R}, k = 1, \cdots, N$ be the $k$th block-maximum, with block size $B_k$, for an underlying strictly stationary process, $s_t$.  That is 
\begin{equation}
y_k = \max_{N_{k-1} < t \leq N_{k}} s_t,
\end{equation}
where $( N_{k-1}, N_k ]$ are the indices for the $k$th block {($k\geq 1$)} and $N_0 = 0$. For our study, $s_t\in\mathbb{R}$ is the daily precipitation recorded at a particular NOAA weather station and $y_k$ is the largest daily precipitation amount over the course of a year. Similar to the central-limit theorem, there is a limiting distribution for the normalized block-maximum of i.i.d. random variables (RVs). The Fisher-Tippet-Gnedenko (FTG) theorem states that the only non-degenerate limiting distribution for the block-maximum, is in the generalized extreme value (GEV) family \citep{ekm_MEE_2003} with shape parameter, or tail-index, $\xi$. That is, the cumulative distribution function (cdf) of the block-maximum, $y_k$, properly normalized, converges to a distribution  $H_{\xi}$ as $(N_k - N_{k-1})  \to \infty$ which has the Jenkinson-von Mises representation \citep{smith2003},

\begin{equation}
\label{GEVcdf}
H_{ \xi }(y) = \exp( - (1+ \xi y )^{-1/\xi} ).
\end{equation}

When the tail-index, $\xi$, is positive, which is our focus, the resulting distribution is the Fr\'{e}chet distribution with $y \geq -1/\xi$.  The Fr\'{e}chet is the limiting distribution for many heavy-tailed underlying random variables \citep{resnick_2007}.  Upon normalization, a three parameter family is obtained and, therefore, we asymptotically model each of the block-maxima $y_k \sim H_{\xi, \beta, \mu} (y_k) = H_{\xi} ( \frac{y_k - \mu}{ \beta})$, with $\xi, \beta > 0$ and support $y_k \in [\mu - \beta/\xi, \infty)$. 

The three parameters of the GEV distribution, $\xi, \beta, \mu$, define its shape, scaling, and location. Given an observation of a block-maximum, $y$, the likelihood function is $p(y|\xi, \beta, \mu) = \frac{d}{d y} H_{\xi} ( \frac{y - \mu}{ \beta})$.  A reasonable, and simplifying, assumption is that the block-maximum, $y$, has support $[0,\infty)$ which allows the location parameter to be absorbed as $\mu = \beta/\xi$ with the implication being that  there is at least one precipitation event each year. The GEV cumulative distribution can then be written as
\begin{equation}
H_{\xi, \beta}(y) = \exp(-(\frac{\xi}{\beta} y)^{-1/\xi}).
\end{equation}

and the likelihood function, or probability density, can be expressed as
\begin{equation}
\label{eq:like}
    p(y|\xi, \beta) = \frac{1}{\beta} (\frac{\xi}{\beta} y)^{-1-1/\xi} \exp(-(\frac{\xi}{\beta} y)^{-1/\xi})
\end{equation}

Under the assumption of independence, the likelihood of the complete set of block maximum observations, $y_{1:N}\equiv \{y_1, y_2, \cdots, y_{N}\}$, is the product of the individual likelihoods,
\begin{equation}
    p(y_{1:N}|\xi, \beta) = \prod_{i=1}^N p(y_i|\xi, \beta).
\end{equation}

One of the main issues in EVA is the trade off between block size and the number of blocks.  The block size needs to be large enough to validate the independence and GEV assumptions but a large number of blocks are needed for parameter estimation.  The benefit of a Bayesian approach is parameter uncertainty is naturally incorporated so more emphasis can be placed on choosing the appropriate block size.

Traditionally, the likelihood function is maximized with respect to the parameters to produce maximum likelihood (ML) estimates $\hat{\xi}_{ML}$ and $\hat{\beta}_{ML}$ and these are then used to make inferences for the future \citep{coles}. Ideally, the goal is to estimate quantile levels, $\eta_{\alpha}$, for levels of $\alpha$ such as 99\% or 99.9\%. These quantile levels, $\eta_{\alpha}$, are also known as the $1/(1-\alpha)$-year return level. For example, a precipitation event of $\eta_{99\%}$ defines the 100-year return level which is often crudely referenced as a one in 100 year extreme precipitation event.

Care must be taken in interpreting GEV return levels as they are quantiles for the maximum event over a period of time, typically a year \citep{coxnorthrop2002}.  For example, the 100-year return level actually has a 63\% probability of occurring at least once over a 100-year period, which may seem surprising. This is because we are using the cdf of the annual maximum daily event as opposed to the cdf of the $N$-year block-maximum, $H^N_{\xi, \beta}(y)$. That said, these predictive quantiles, or return-levels, are a vital yardstick used by policy makers, engineers, and risk managers. 

Formally, the ML-quantile estimate is computed by inverting the GEV distribution using the ML estimates,  

\begin{equation}
\hat{\eta}_{\alpha} = H^{-1}_{\hat{\xi}_{ML}, \hat{\beta}_{ML}}(\alpha) = \frac{\hat{\beta}_{ML}}{\hat{\xi}_{ML}} (-\ln(\alpha))^{-\hat{\xi}_{ML}}.
\end{equation}

The main issue with the maximum likelihood approach is that one single value for the parameters are used.  This compares with a Bayesian methodology which integrates, or marginalizes, over all possible parameter values, weighted by the posterior given the observations. In other words, the return-level, $H^{-1}_{\xi, \beta}(\alpha)$ is considered a random variable from which estimates are derived.  For example,
\begin{equation}
    E(\eta_{\alpha}) = E_{\xi, \beta | y_{1:N}}( H^{-1}_{\xi, \beta}(\alpha) )
\end{equation}
where the expectation is taken over the posterior density 
\begin{equation}
\label{poster}
    p(\xi, \beta | y_{1:N} ) \propto  p(y_{1:N}|\xi, \beta)  p(\xi, \beta)
\end{equation}
which, in the case of flat priors, is the likelihood normalized to unity.  Other statistics of possible interest would be the median return-level, the maximum a posteriori (MAP) return-level, and percentiles of the return-level distribution. Marginal distributions, such as  $p(\xi | y_{1:N} )$, can be determined by integrating (\ref{poster}) over a subset of the parameters.

While the Bayesian and maximum likelihood methods generally result in similar forecasts when there are a large number of observations, the difference for smaller data sets can be significant.  This is particularly true for probabilistic queries deep in the tail of the distribution. This is due to the impact of parameter uncertainty which leads to higher estimated Bayesian return-levels than when ML estimates are employed \citep{smith2003}. As we will see in the sequel, this is important for climate data as data sets are usually limited in size or there is a need to account for nonstationarity by using smaller time series lengths \citep{atg_cmc_2018}.

\section{Results}
\label{sec:results}

\subsection{Stationary Analysis: 1938-2021}

To illustrate the performance of the Bayesian approach, we first performed the analysis on the complete set of data shown in Figure \ref{fig:fig1}, which consisted of 84 years (1938-2021) with each year's datum being the maximum daily precipitation. The likelihood function was computed over a grid of possible values for $\xi$ and $\beta$. This was computationally feasible given our reduced parameter set (i.e., $\mu = \beta/\xi$). The maximum likelihood estimates for the GEV parameters were $\hat{\xi}_{ML} = .3176$ and $\hat{\beta}_{ML} = .7833$. These resulted in estimates for the $N$-year return levels shown in the first column of Table \ref{tab:table1}. The ML estimate for the 100-year return level ($\hat{\eta}_{99\%}$) is 10.63 inches which policy makers might consider as an estimate of the daily precipitation amount that can be expected once over a 100-year period \footnote{As noted previously, this is not exactly a true statement since we are dealing with annual maximums. For example, using the ML estimates, the daily precipitation amount that results in a 50\% chance of non-occurrence would be 11.96 inches. That said, in some applications, where an exceedence leads to critical failure, the point may be moot.}.

\begin{table}[h]
	\caption{GEV Parameter and Return Level Estimates}
	\centering
	\begin{tabular}{l || c |cccc}
		\toprule
		     & ML     & Mean & Median & 5\% & 95\% \\
		\midrule
		$\hat{\xi}$ & .3176 & .3279 &  .3250 & .2840 & .3760 \\
		$\hat{\beta}$ & .7833 & .8126 &  .8040 & .6820 & .9620 \\
		\midrule
		$N = 10\;\; (\eta_{90\%})$  & 5.04  & 5.20   & 5.17  & 4.56  & 5.99\\ 
		$N = 25 \;\; (\eta_{96\%})$  & 6.81  & 7.12  & 7.04  & 6.00  & 8.54\\ 
		$N = 100 \;\; (\eta_{99\%})$ & 10.63 & 11.34  & 11.12 & 8.97  & 14.47\\ 
		$N = 500 \;\; (\eta_{99.8\%})$ & 17.75 & 19.41  & 18.84 & 14.24 & 26.53\\ 
		\bottomrule
	\end{tabular}
	\label{tab:table1}
\end{table}

\begin{figure}[h]
	\centering
    \includegraphics[width=.65\textwidth]{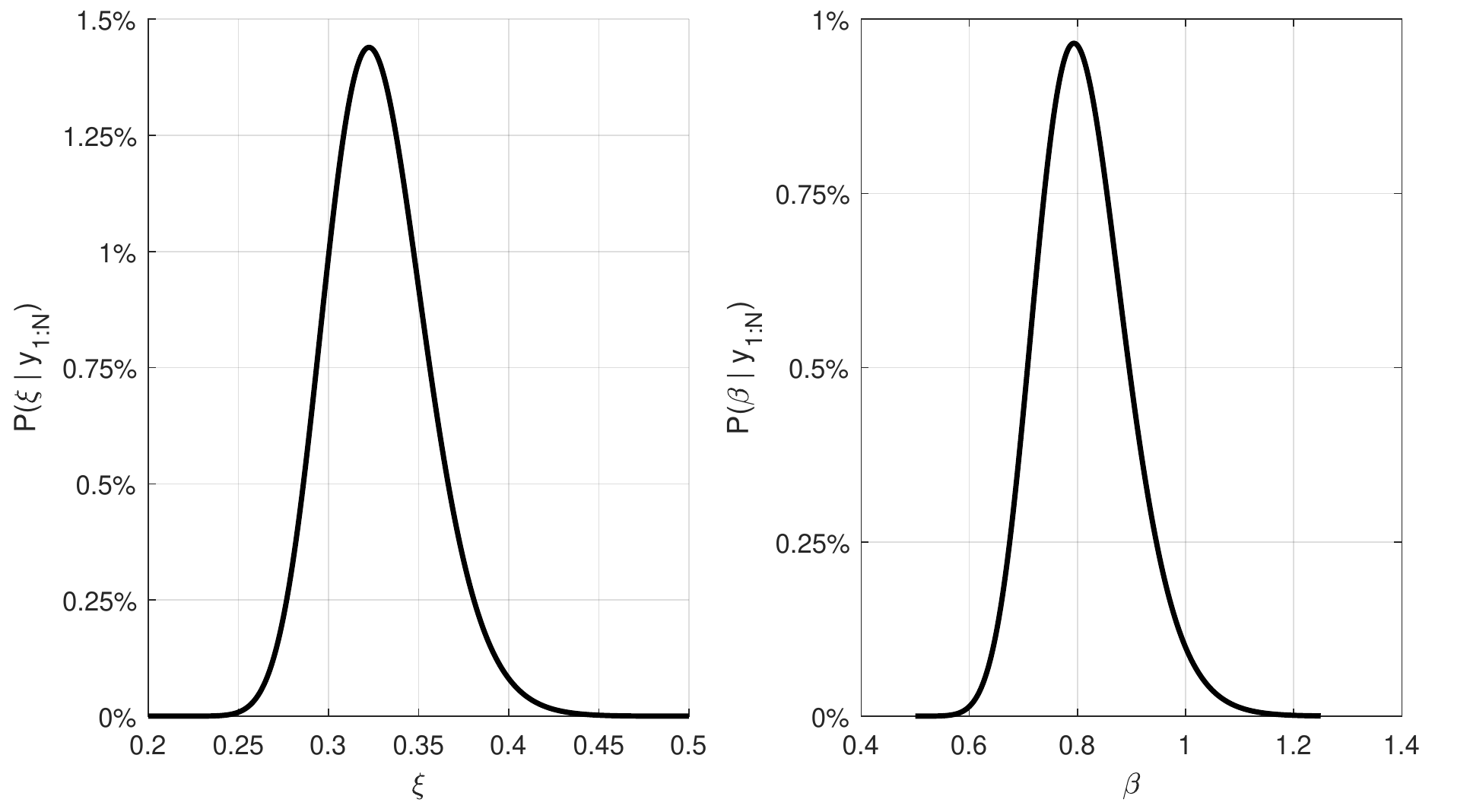}
    \caption{Marginal Probability Density Functions}	
    \label{fig:fig2}
\end{figure}

Shown in Figure \ref{fig:fig2} are the derived marginal probability density functions for the two GEV parameters and Bayesian estimates for the parameters are shown in Table \ref{tab:table1}.  There is strong evidence that the tail-index $\xi> .25$, which indicates the underlying daily rainfall random variable is from the heavy-tail distribution class, with power-law tail behaviour, and that fourth order moments and higher do not exist. The marginal distributions illustrate the wide range of possible parameter values with, for example, a 90\% credible interval (CI) of [.284, .376] for $\xi$ and [.682, .962] for $\beta$.  There is also a right-skew to the distributions, which factor into return-level forecasts.  

\begin{figure}[h]
	\centering
    \includegraphics[width=.5\textwidth]{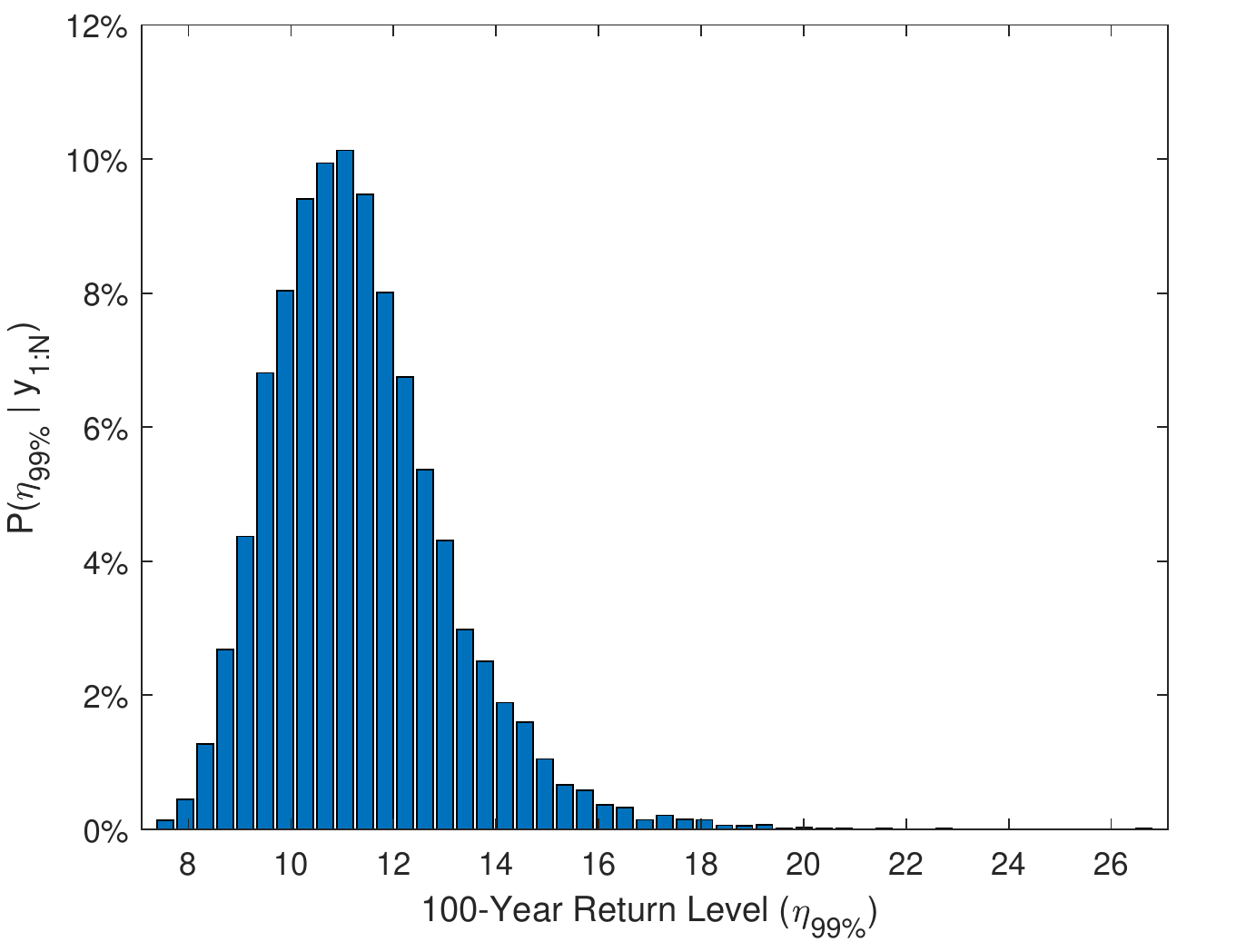}
    \caption{Frequency Distribution for 100-year Return Level}	
    \label{fig:fig3}
\end{figure}

What is not shown in Figure \ref{fig:fig2} is the high correlation, of .94, between the parameters which results in a more skewed return-level distribution.  For example, shown in Figure \ref{fig:fig3} is the frequency distribution of 100-year return levels derived from a bootstrap method applied to the GEV parameters. Specifically, 10000 random samples for $[ \xi, \beta]$ were generated from the posterior (\ref{poster}) and used to compute $\eta_{99\%}$, the frequency distribution of which is shown in Figure \ref{fig:fig3}. The mean of this distribution, which is also given in Table \ref{tab:table1} is 11.34 inches with a 95\% CI of [8.97, 14,47] and there is a significant right skew of 5. 

This analysis shows that the Bayes method results in higher predicted precipitation risk than the ML method. For example, 100-year return levels are .5-1 inch higher using Bayes estimates.  More significantly, the mean 100-year return level that results in a 50\% chance of non-occurrence would be an even higher 12.83 inches, more than 2 inches above the traditional estimate.  This highlights the difference in assessed precipitation risk between the Bayesian and ML methods and that the difference increases when considering the cdf of the $N$-year block-maximum, $H^N_{\xi, \beta}(y)$. 

\subsection{Non-Stationary Analysis: 1938-1988 vs. 1989-2021}

Non-stationary analysis of climate data is of increasing importance \citep{IPCC2014}.  with the main challenge being either the necessity of using smaller time windows \citep{atg_cmc_2018} or modeling the time dependence either directly as, for example, a trend \citep{Cheng_2014} or in a dynamic state-space model \citep{dej_2021}.

\begin{figure}[h]
	\centering
    \includegraphics[width=.5\textwidth]{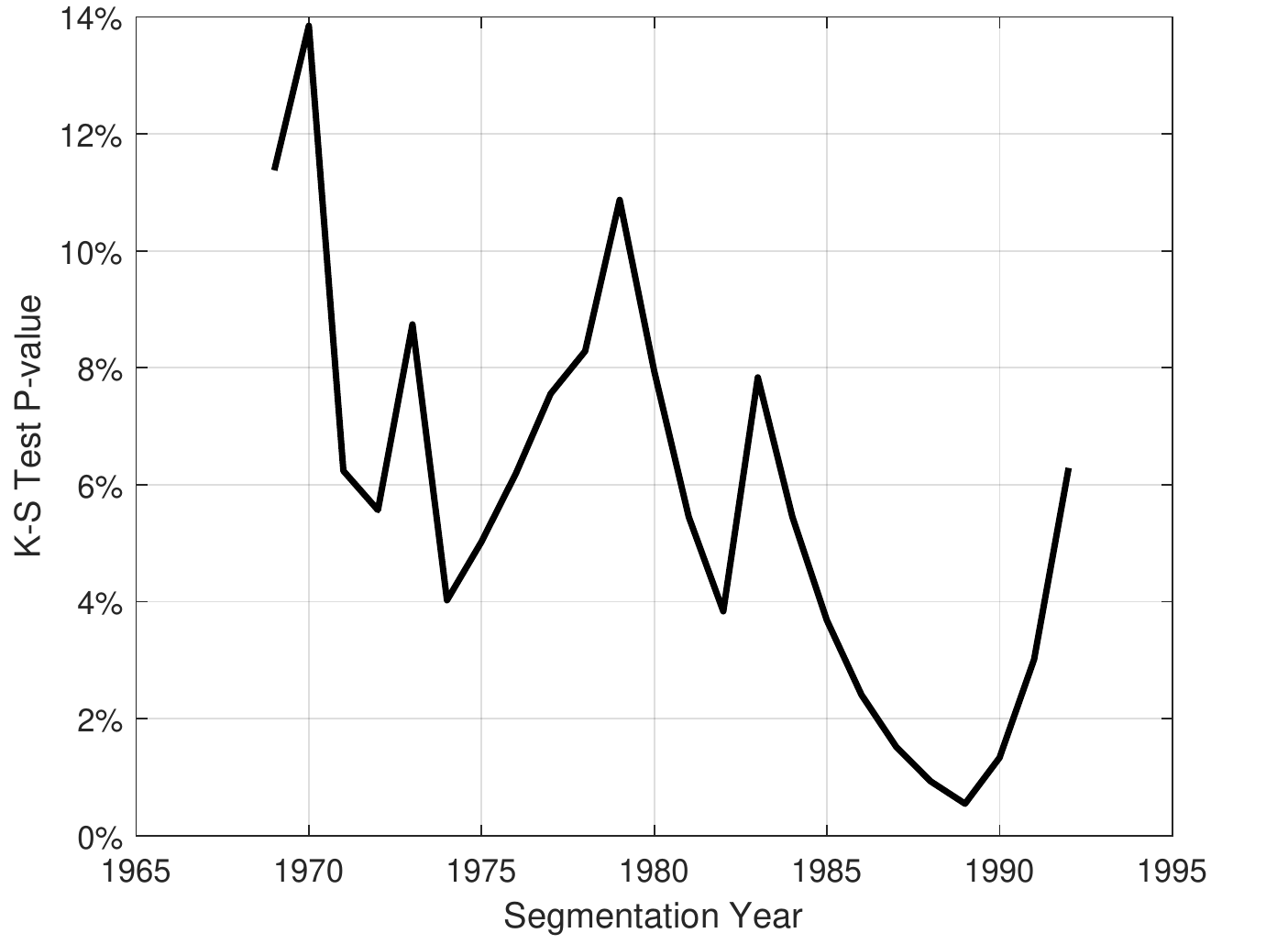}
    \caption{Kolmogorov-Smirnov Test}	
    \label{fig:fig4}
\end{figure}

We applied a Kolmogorov-Smirnov test to the data, segmenting the data into non-overlapping time periods with a minimum of 30 years to each segment. The results are shown in Figure \ref{fig:fig4} which indicates a segmentation break in the data with the lowest p-value of .005 implying there is substantial evidence that the data from 1938-1988 is from a different distribution than from 1989-2021 and that the latter time period has a larger tail.  A basic two sample t-test confirms this result with a p-value of .02 although the underlying assumption of normality is questionable. The Mann-Kendall trend test is inconclusive for those two separate time periods but for the complete set of data a p-value of .05 suggests the potential need for a dynamic model for the GEV parameters as in \citep{dej_2021}. 

\begin{figure}[h]
	\centering
    \includegraphics[width=.5\textwidth]{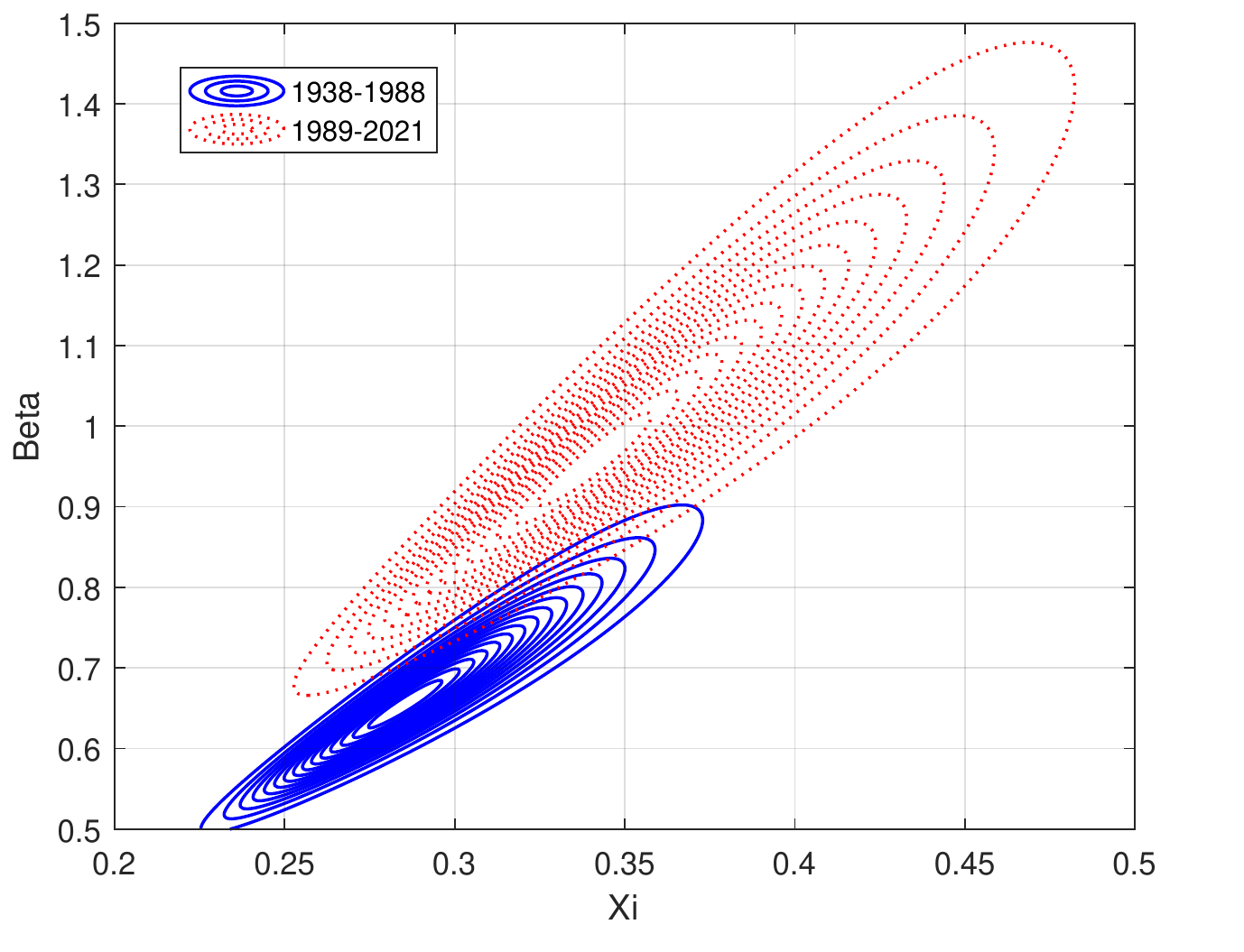}
    \caption{Posterior $p(\xi, \beta | y_{M:N} )$ }	
    \label{fig:fig5}
\end{figure}

Our goal in this paper is not to model the apparent non-stationarity in the data, for which a dynamic Bayesian model is ideal, but rather to show the efficacy of the Bayesian methodology under small sample sizes. We ran our stationary analysis on the two windows of data, 1938-1988 and 1989-2021, and the results highlight a significant change in the extreme precipitation data.  Shown in Figure \ref{fig:fig5} is a contour plot of the posterior density, or likelihood function, for the GEV parameters. We can see a number of features including the correlation in the parameters, asymmetry of the density, and the dispersion, which is more substantial for the smaller time window (1989-2021) where there are only 33 samples.

\begin{table}[h]
	\caption{GEV Parameter and 100-Year Return Level Estimates}
	\centering
	\begin{tabular}{c || cc | cccc | ccccc }
		\toprule
		 & &&&&&& \multicolumn{5}{c}{$\hat{\eta}_{99\%}$} \\ 
		     & $\hat{\mu}$ & $\hat{\sigma}$ & $\hat{\xi}_{ML}$ & $\hat{\beta}_{ML}$ & $\hat{\xi}_{50\%}$ & $\hat{\beta}_{50\%}$ & ML & Median & Mean & 5\%-tile & 95\%-tile  \\
		\midrule
		1938-2021 & 3.21 & 1.62 & .318 & .783 & .325 & .804 & 10.63 & 11.12 & 11.34 & 8.97 & 14.47 \\
		\midrule
		1938-1988 & 2.84 & 1.03 & .285 & .654 & .297 & .684 & 8.52 & 9.09 & 9.38 & 7.12 & 12.53 \\
		\midrule
        1989-2021 & 3.78 & 2.14 & .341 & .956 & .364 & 1.02 & 13.45 & 15.25 & 16.28 & 10.52 & 25.65 \\
		\bottomrule
	\end{tabular}
	\label{tab:table2}
\end{table}

The estimates for the GEV parameters and the 100-year return level for the whole period and the two time periods is shown in Table \ref{tab:table2}.  We also included the sample mean and standard deviation of the block-maxima. We can see the large increase in the estimated GEV parameters and return levels between the two time periods. The increase in return-level estimates range from 5 inches, for the ML, to 7 inches for the mean estimates. In addition, estimates for the latter time period are well outside the 90\% credible intervals of the former, and vice-versa.

\begin{figure}[h]
	\centering
    \includegraphics[width=.5\textwidth]{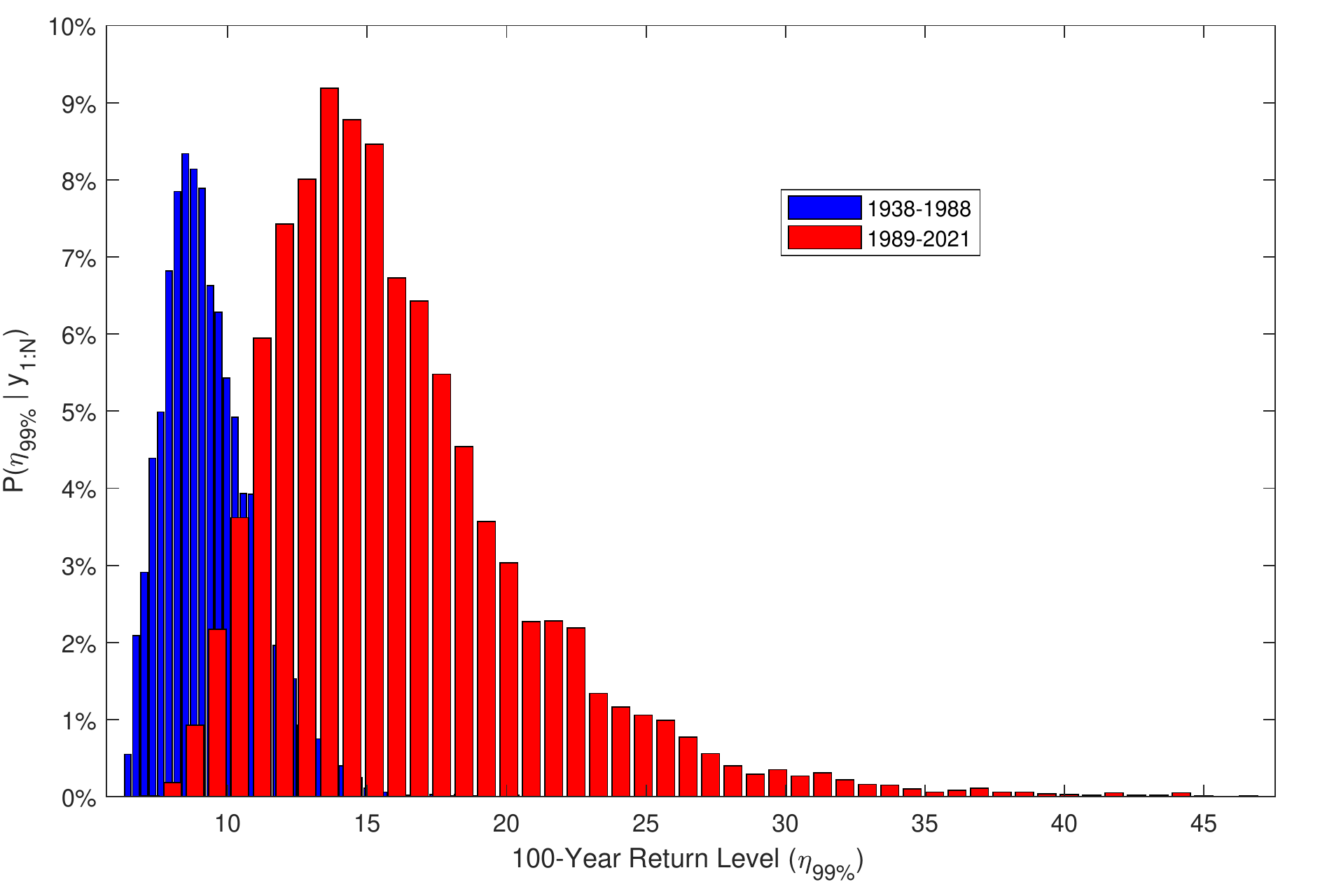}
    \caption{Frequency Distributions, derived from Monte-Carlo Sampling, for 100-Year Return Level}	
    \label{fig:fig6}
\end{figure}

\begin{figure}[h]
	\centering
    \includegraphics[width=.5\textwidth]{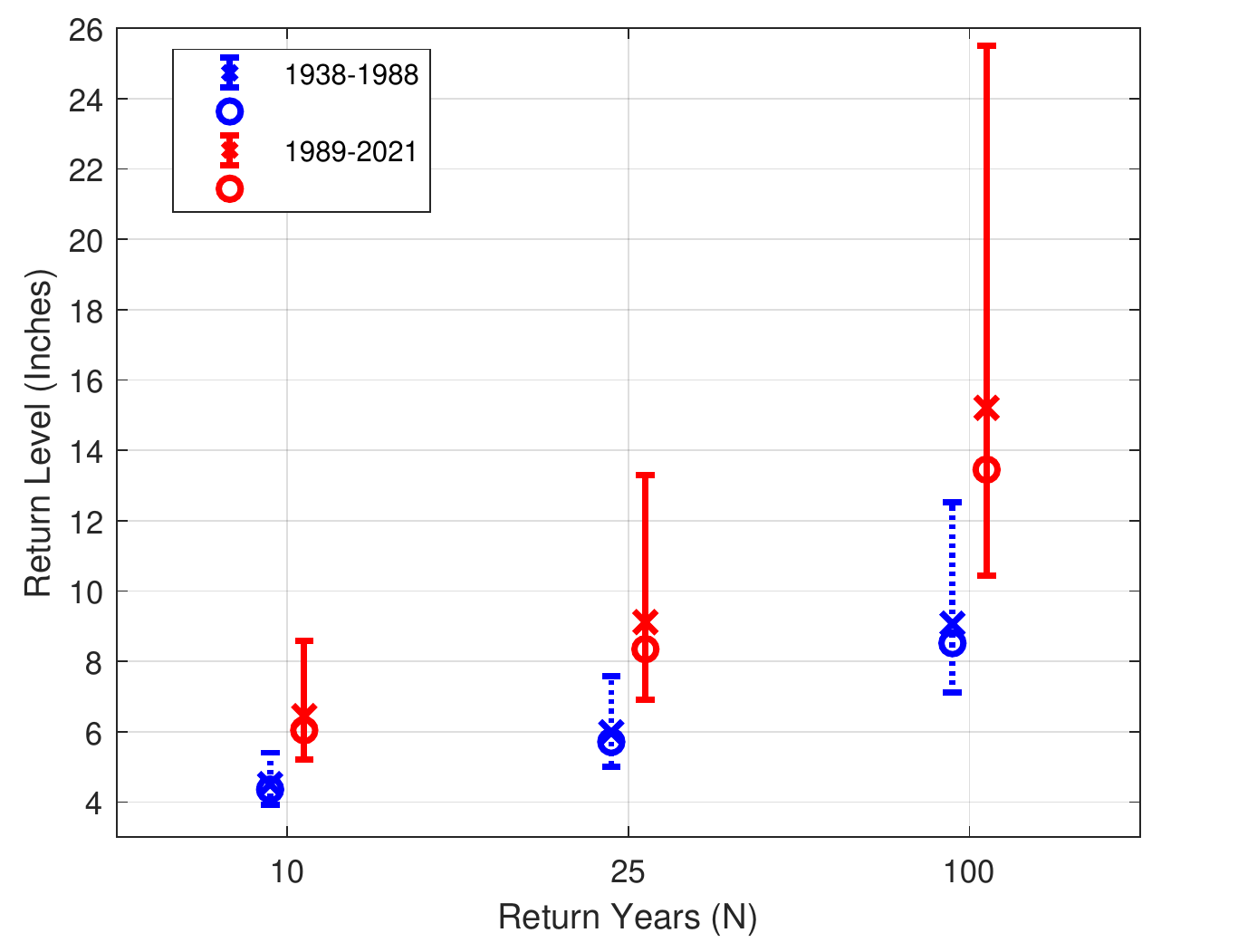}
    \caption{Statistics for Return Level for Different Return Years (ML estimate 'O', Median Estimate 'X', and 90\% credible interval (5-95\%).}	
    \label{fig:fig7}
\end{figure}

The frequency distribution for the 100-year return level, generated from sampling the posterior, is shown Figure \ref{fig:fig6} where we can visualize the dramatic shift in the posterior due to the change in observed precipitation data. Randomly sampling from these distributions strongly supports this claim as there is a 96\% chance that a sample generated from the 1989-2021 data is greater than one from 1938-1988 and there is only a 21\% chance that either cohort's sample lies in the 90\% CI of the others. Lastly, we highlight the difference in return-levels forecasts in Figure \ref{fig:fig7} where we plot the ML ('O') and median ('X') return level estimates and the 90\% credible interval for the 10, 25, and 100-year return levels. We can see the small overlap of confidence intervals. This highlights the structural shift in the data.

One thing we note from the data is the large sample that occurred in August 2014, where there was a recorded 13.5 inches of rain in one day at Islip, NY. This eclipsed the previous NY record of 11.6 inches and was a highly localized event.  At nearby Farmingdale, NY Republic airport (about 20 miles away), the recorded maximum for 2014 was only 5 inches. Given this extreme event in the data, one might wonder its impact on the results. For purposes of illustration, we modified the 2014 data and reran the analysis.  While different, the broad results remain intact.  For example, the median 100-year return estimate declined from 15.25 to 13.61 inches but this was still outside the 90\% credible interval for the prior period.

\section{Conclusion}
The Bayesian methodology for data analysis offers significant advantages to its frequentist counterpart (e.g., ML).  The most important being that the complete posterior of the distribution, given the observed data, is used rather than relying on a single point estimate. We employed a Bayesian approach to analysing extreme precipitation data illustrating that significant differences can arise due to the nature of the likelihood function.  This is particularly true for small sample sizes which are regularly encountered in climate data, particularly when non-stationarity (i.e., climate change) is addressed. As we showed, the Bayesian approach resulted in higher levels of precipitation risk than stated by the ML method and that the difference was larger when assessing lower probability events farther into the future.  Given policy makers should be risk-averse by nature, the Bayesian method appears to offer a more complete assessment of risk that policy makers can rely on.

\bibliographystyle{unsrtnat}
\bibliography{DEJRefs}  

\begin{thebibliography}{19}
\providecommand{\natexlab}[1]{#1}
\providecommand{\url}[1]{\texttt{#1}}
\expandafter\ifx\csname urlstyle\endcsname\relax
  \providecommand{\doi}[1]{doi: #1}\else
  \providecommand{\doi}{doi: \begingroup \urlstyle{rm}\Url}\fi

\bibitem[Broadwater and Chellappa(2010)]{broadwater2010}
J.~B. Broadwater and R.~Chellappa.
\newblock Adaptive threshold estimation via extreme value theory.
\newblock \emph{IEEE Transactions on Signal Processing}, 58\penalty0
  (2):\penalty0 490--500, Feb 2010.

\bibitem[Resnick and Rootze\'n(2000)]{ResRootz2000}
S.~I. Resnick and H~Rootze\'n.
\newblock {Self-similar} communication models and very heavy tails.
\newblock \emph{Ann. Appl. Probab.}, 10\penalty0 (3):\penalty0 753--778, August
  2000.

\bibitem[Roberts(2000)]{SJRoberts2000}
S.~J. Roberts.
\newblock Extreme value statistics for novelty detection in biomedical data
  processing.
\newblock \emph{IEE Proc. Sci., Meas. Technol.}, 147\penalty0 (6):\penalty0
  363--367, November 2000.

\bibitem[Shenoy and Gorinevsky(2015)]{shenoy2015}
S.~Shenoy and D.~Gorinevsky.
\newblock Estimating long tail models for risk trends.
\newblock \emph{IEEE Signal Processing Letters}, 22\penalty0 (7):\penalty0 968
  -- 972, July 2015.

\bibitem[Anderson et~al.(2017)Anderson, Branch, Cooper, and
  Dulvy]{Anderson3252}
S.~C. Anderson, Trevor~A. Branch, Andrew~B. Cooper, and Nicholas~K. Dulvy.
\newblock Black-swan events in animal populations.
\newblock \emph{Proceedings of the National Academy of Sciences}, 114\penalty0
  (12):\penalty0 3252--3257, 2017.

\bibitem[Johnston and Djuri\'c(2011)]{dej_pmd_SPM2011}
D.~E. Johnston and P.~M. Djuri\'c.
\newblock The science behind risk management: A signal processing perspective.
\newblock \emph{Signal Process. Mag.}, 28\penalty0 (5):\penalty0 26--36,
  September 2011.

\bibitem[DeGaetano and Castellano(2017)]{atg_cmc_2017}
A.~T. DeGaetano and C.~M. Castellano.
\newblock Future projections of extreme precipitation
  intensity-duration-frequency curves for climate adaption planning in {New
  York} state.
\newblock \emph{Climate Services}, 5:\penalty0 23--35, January 2017.

\bibitem[Embrechts et~al.(2003)Embrechts, Kluppelberg, and
  Mikosch]{ekm_MEE_2003}
P.~Embrechts, C.~Kluppelberg, and T.~Mikosch.
\newblock \emph{Modelling Extremal Events}.
\newblock Springer-Verlag, New York, 2003.

\bibitem[Cox et~al.(2002)Cox, Isham, and Northrop]{coxnorthrop2002}
D.~R. Cox, V.~S. Isham, and P.~J. Northrop.
\newblock Floods: some probabilistic and statistical approaches.
\newblock \emph{Phil. Trans. R. Soc. Lond. A}, 360\penalty0 (1):\penalty0
  1389--1408, July 2002.

\bibitem[Coles and Dixon(1999)]{coles_1999}
S.~G. Coles and M.~J. Dixon.
\newblock Likelihood-based inference for extreme value models.
\newblock \emph{Extremes}, 2\penalty0 (1):\penalty0 5--23, March 1999.

\bibitem[Nerantzaki and Papalexiou(2022)]{Nerantzaki_2022}
S.~D. Nerantzaki and S.~M. Papalexiou.
\newblock Assessing extremes in hydroclimatology: A review on probabilistic
  methods.
\newblock \emph{Journal of Hydrology}, 605:\penalty0 1--20, February 2022.

\bibitem[Coles and Powell(1996)]{coles_1996}
S.~G. Coles and E.~A. Powell.
\newblock Bayesian methods in extreme value modelling: A review and new
  developments.
\newblock \emph{International Statistical Review}, 64\penalty0 (1):\penalty0
  119--136, April 1996.

\bibitem[Smith(2003)]{smith2003}
R.~L. Smith.
\newblock Statistics of extremes with applications in environment, insurance
  and finance.
\newblock In B.~Finkenst{\"a}dt and H.~Rootze\'n, editors, \emph{Extreme Values
  in Finance, Telecommunications and the Environment}, pages 1--78. Chapman and
  Hall CRC, 2003.

\bibitem[Resnick(2008)]{resnick_2007}
S.~Resnick.
\newblock \emph{Heavy Tail Phenomena: Probabilistic and Statistical Modeling}.
\newblock Springer-Verlag, New York, 2008.

\bibitem[Coles(2004)]{coles}
S.~Coles.
\newblock \emph{An Introduction to Statistical Modeling of Extreme Values}.
\newblock Springer-Verlag, London, UK, 2004.

\bibitem[DeGaetano and Castellano(2018)]{atg_cmc_2018}
A.~T. DeGaetano and C.~M. Castellano.
\newblock Selecting time series length to moderate the impact of
  nonstationarity in extreme rainfall analysis.
\newblock \emph{Journal of Applied Meteorology and Climatology}, 57\penalty0
  (10):\penalty0 2285--2296, October 2018.

\bibitem[IPCC(2014)]{IPCC2014}
IPCC.
\newblock In Core~Writing Team, R.~K. Pachauri, and L.~A. Myer, editors,
  \emph{Synthesis Report: Contribution of Working Groups I, II, and III to the
  Fifth Assessment Report of the Intergovernmental Panel on Climate Change},
  page 151. IPCC, 2014.

\bibitem[Cheng et~al.(2014)Cheng, AghaKouchak, Gilleland, and Katz]{Cheng_2014}
L.~Cheng, A.~AghaKouchak, E.~Gilleland, and R.~W. Katz.
\newblock Non-stationary extreme value analysis in a changing climate.
\newblock \emph{Climate Change}, 127:\penalty0 353--369, September 2014.

\bibitem[Johnston(2021)]{dej_2021}
D.~E. Johnston.
\newblock Bayesian forecasting of dynamic extreme quantiles.
\newblock \emph{Forecasting}, 3\penalty0 (4):\penalty0 729--740, October 2021.

\end{thebibliography}






\end{document}